%% file: TED.tex
\documentclass[11pt]{llncs}

\usepackage{comment}

\usepackage{geometry}%
\usepackage{amsmath}
\usepackage{epsfig}%
\usepackage{graphics}%
\usepackage{multicol}
\usepackage{enumerate}

\newcommand{\height}{\mr{height}}

\newcommand{\leaves}{\mr{leaves}}
\newcommand{\mr}{\mathrm}

\newcommand{\cmatch}{c_{{}_{\mr{{rel}}}}}
\newcommand{\cdel}{c_{{}_{\mr{{del}}}}}

\newcommand{\Left}{\mathsf{left}}
\newcommand{\Right}{\mathsf{right}}
\newcommand{\tu}{\textup}

\newcommand{\TopLight}{\mathrm{TopLight}}
\newcommand{\ldepth}{\mathrm{ldepth}}

\newcommand{\light}{\mathrm{light}}

\begin{document}

\title{An $O(n^3)$-Time Algorithm for Tree Edit Distance}

\author{Erik D. Demaine \inst{} \and Shay Mozes\thanks{Work conducted while visiting
MIT}\inst{} \and Benjamin Rossman\inst{} \and Oren Weimann\inst{}}

\institute{MIT Computer Science and Artificial Intelligence
Laboratory,\\ 32 Vassar Street, Cambridge, MA 02139, USA.\\
\email{edemaine@mit.edu,shaymozes@gmail.com,brossman@mit.edu,oweimann@mit.edu}}

\begin{twocolumn}
\date{}
\maketitle \pagestyle{plain}
\begin{abstract}
The {\em edit distance} between two ordered trees with vertex labels
is the minimum cost of transforming one tree into the other by a
sequence of elementary operations consisting of deleting and
relabeling existing nodes, as well as inserting new nodes. In this
paper, we present a worst-case $O(n^3)$-time algorithm for this
problem, improving the previous best $O(n^3\log n)$-time
algorithm~\cite{Klein}. Our result requires a novel adaptive
strategy for deciding how a dynamic program divides into subproblems
(which is interesting in its own right), together with a deeper
understanding of the previous algorithms for the problem. We also
prove the optimality of our algorithm among the family of
\emph{decomposition strategy} algorithms---which also includes the
previous fastest algorithms---by tightening the known lower bound of
$\Omega(n^2\log^2 n)$~\cite{Touzet} to $\Omega(n^3)$, matching our
algorithm's running time. Furthermore, we obtain matching upper and
lower bounds of $\Theta(n m^2 (1 + \log \frac{n}{m}))$ when the two
trees have different sizes $m$ and~$n$, where $m < n$.
\end{abstract}

\section{Introduction}
\input{Introduction}

\section{Background and Framework \label{pre}}
\input{Preliminaries}

\section{The Algorithm \label{algo}}
\input{Algorithm}

\section{A Tight Lower Bound for Strategy Algorithms\label{lowerbound}}
\input{LowerBound}

\section{Conclusions \label{conclusions}}
\input{Conclusions}

\Large
\bibliographystyle{plain}
\bibliography{TED}

\end{twocolumn}
\end{document}

%% file: Introduction.tex

The problem of comparing trees occurs in diverse areas such as
structured text databases like XML, computer vision,
compiler optimization, natural language processing, and
computational biology~\cite{Bille,XML,KleinTirthapuraSODA2000,Shasha,Tai}.

\begin{figure}[h!]
\begin{center}
\includegraphics[scale=0.7]{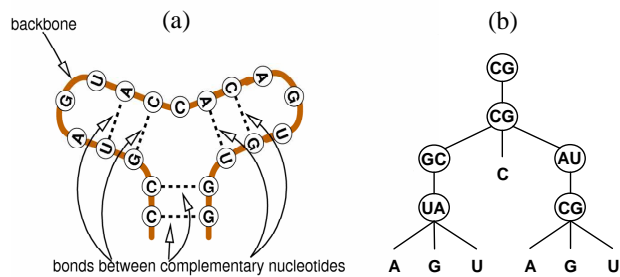}
\caption{\label{RNA} Two different ways of viewing an RNA sequence.
In (a), a schematic 2-dimensional description of an RNA folding. In
(b), the RNA as a rooted ordered tree.}
\end{center}
\end{figure}

As an example, we describe an application in computational biology.
\emph{Ribonucleic acid} (RNA) is a polymer consisting of a sequence
of nucleotides (Adenine, Cytosine, Guanine, and Uracil) connected
linearly via a backbone. In addition, complementary nucleotides
(A--U, G--C, and G--U) can form hydrogen bonds, leading to a
structural formation called the \emph{secondary structure} of the
RNA. Because of the nested nature of these hydrogen bonds, the
secondary structure of RNA can be represented by a rooted ordered
tree, as shown in Fig.~\ref{RNA}. Recently, comparing RNA sequences
has gained increasing interest thanks to numerous discoveries of
biological functions associated with RNA. A major fraction of RNA's
function is determined by its secondary structure~\cite{Moore:99}.
Therefore, computing the similarity between the secondary structure
of two RNA molecules can help determine the functional similarities
of these molecules.

\begin{figure}[h!]
\centerline{
\includegraphics[scale=0.52]
{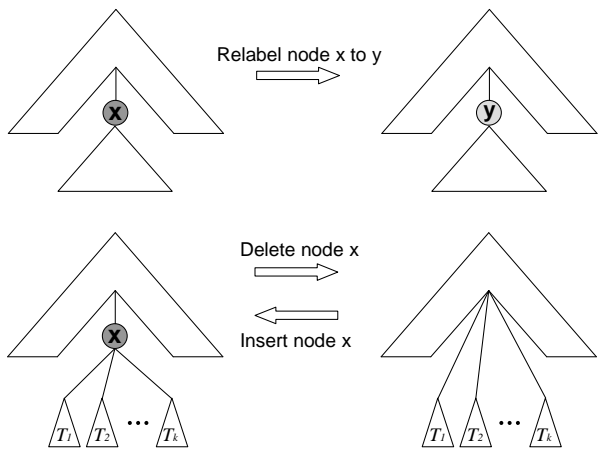} } \caption{The three editing operations on a
tree with vertex labels.} \label{Operations}
\end{figure}

The {\em tree edit distance} metric is a common similarity measure
for ordered trees, introduced by Tai in the late 1970's~\cite{Tai}
as a generalization of the well-known string edit distance
problem~\cite{StringED}. Let $F$ and $G$ be two rooted trees with a
left-to-right order among siblings and where each vertex is assigned
a label from an alphabet $\Sigma$. The \emph{edit distance} between
$F$ and $G$ is the minimum cost of transforming $F$ into $G$ by a
sequence of elementary operations consisting of deleting and
relabeling existing nodes, as well as inserting new nodes (allowing
at most one operation to be performed on each node). These
operations are illustrated in Fig.~\ref{Operations}. The cost of
elementary operations is given by two functions, $\cdel$ and
$\cmatch$, where $\cdel(\tau)$ is the cost of deleting or inserting
a vertex with label $\tau$, and $\cmatch(\tau_1,\tau_2)$ is the cost
of changing the label of a vertex from $\tau_1$ to $\tau_2$. A
deletion in $F$ is equivalent to an insertion in $G$ and vice versa,
so we can focus on finding the minimum cost of a sequence of
deletions and relabels in both trees that transform $F$ and $G$ into
isomorphic trees.

\paragraph{Previous results.}
To state running times, we need some basic notation. Let $n$ and $m$
denote the sizes $|F|$ and $|G|$ of the two input trees, ordered so
that $n \geq m$. Let $n_\leaves$ and $m_\leaves$ denote the
corresponding number of leaves in each tree, and let $n_\height$ and
$m_\height$ denote the corresponding height of each tree, which can
be as large as $n$ and $m$ respectively.

Tai \cite{Tai} presented the first algorithm for computing tree edit
distance, which requires $O(n_\leaves^2 m_\leaves^2 n m)$ time and
space. Tai's algorithm thus has a worst-case running time of $O(n^3
m^3)=O(n^6)$. Shasha and Zhang~\cite{Shasha} improved this result to
an $O(\min\{n_\height,n_\leaves\} \cdot \min\{m_\height, m_\leaves\}
\cdot n m)$ time algorithm using $O(n m)$ space. In the worst case,
their algorithm runs in $O(n^2 m^2) = O(n^4)$ time.
Klein~\cite{Klein} improved this result to a worst-case $O(m^2n
\log n)=O(n^3\log n)$-time algorithm that uses $O(n m)$ space. In
addition, Klein's algorithm can be adapted to solve an unrooted
version of the problem. These last two algorithms are based on
closely related dynamic programs, and both present different ways of
computing only a subset of a larger dynamic program table; these
entries are referred to as \emph{relevant subproblems}.
In~\cite{Touzet}, Dulucq and Touzet introduced the notion of a
\emph{decomposition strategy} (see Section~\ref{strategy}) as a
general framework for algorithms that use this type of dynamic
program, and proved a lower bound of $\Omega(n m \log n \log m)$
time for any such strategy.

Many other solutions have been developed;
see~\cite{Bille,ValienteBook} for surveys. The most recent
development is by Chen~\cite{Chen}, who presented a different
approach that uses results on fast matrix multiplication. Chen's
algorithm uses $O(n m + n m_\leaves^2 + n_\leaves m_\leaves^{2.5})$
time and $O(n + (m + n_\leaves^2) \min\{n_\leaves,n_\height\})$
space. In the worst case, this algorithm runs in $O(n m^{2.5}) =
O(n^{3.5})$ time. In general, Klein's algorithm remained the best in
terms of worst-case time complexity.

\paragraph{Our results.}
In this paper, we present a new algorithm for tree edit distance
that falls into the same \emph{decomposition strategy} framework
of~\cite{Klein,Shasha,Touzet}. Our algorithm runs in $O(n m^2 (1 +
\log\frac{n}{m})) = O(n^3)$ worst-case time and $O(n m)$ space, and
can be adapted for the case where the trees are not rooted. The
corresponding edit script can easily be obtained within the same
time and space bounds. We therefore improve upon all known
algorithms in the worst-case time complexity. Our approach is based
on Klein's, but whereas the recursion scheme in Klein's algorithm is
determined by just one of the two input trees, in our algorithm the
recursion depends alternately on both trees. Furthermore, we prove a
worst-case lower bound of $\Omega(n m^2 (1 + \log\frac{n}{m}))$ time
on all decomposition strategy algorithms. This bound improves the
previous best lower bound of $\Omega(n m \log n \log m)$
time~\cite{Touzet}, and establishes the optimality of our algorithm
among all decomposition strategy algorithms. Our algorithm is
simple, making it easy to implement, but both the upper and lower
bound proofs require complicated analysis.

\paragraph{Roadmap.}
In Section~\ref{pre} we give simple and unified presentations of the
two well-known tree edit algorithms, on which our algorithm is
based, and the class of decomposition strategy algorithms. We
present and analyze our algorithm in Section~\ref{algo}, and prove
the matching lower bound in Section~\ref{lowerbound}. We conclude in
section~\ref{conclusions}.

%% file: Preliminaries.tex

Both the existing algorithms and ours compute the edit distance of
finite ordered $\Sigma$-labeled forests, henceforth {\em forests}.
The unique empty forest/tree is denoted by $\emptyset$. The vertex
set of a forest $F$ is written simply as $F$, as when we speak of a
vertex $v \in F$. For any forest $F$ and $v \in F$, $\sigma(v)$
denotes the $\Sigma$-label of $v$, $F_v$ denotes the subtree of $F$
rooted at $v$, and $F-v$ denotes the forest obtained from $F$ after
deleting $v$. The leftmost and rightmost trees of $F$ are denoted by
$L_F$ and $R_F$ and their roots by $\ell_F$ and $r_F$. We denote by
$F-L_F$ the forest obtained from $F$ after deleting the entire
leftmost tree $L_F$; similarly $F-R_F$. A forest obtained from $F$
by a sequence of any number of deletions of the leftmost and
rightmost roots is called a \emph{subforest} of $F$.

Given forests $F$ and $G$ and vertices $v \in F$ and $w \in G$, we
write $\cdel(v)$ instead of $\cdel(\sigma(v))$ for the cost of
deleting or inserting $v$, and we write $\cmatch(v,w)$ instead of
$\cmatch(\sigma(v),\sigma(w))$ for the cost relabeling $v$ to $w$.
$\delta(F,G)$ denotes the edit distance between the forests $F$ and
$G$.

Because insertion and deletion costs are the same (for a node of a
given label), {\em insertion in one forest} is tantamount to {\em
deletion in the other forest}. Therefore, the only edit operations
we need to consider are relabels and deletions of nodes in both
forests. In the next two sections, we briefly present the algorithms
of Shasha and Zhang, and of Klein. Our presentation is inspired by
the tree similarity survey of Bille~\cite{Bille}, and is essential
for understanding our algorithm.

\subsection{Shasha and Zhang's Algorithm~\cite{Shasha}}
\label{Zhang} Given two forests $F$ and $G$ of sizes $n$ and $m$
respectively, the following lemma is easy to verify. Intuitively,
this lemma says that the two rightmost roots in $F$ and $G$ are
either matched with each other or one of them is deleted.
\begin{lemma}[\cite{Shasha}] \label{shasha}
$\delta(F,G)$ can be computed as follows:
\begin{enumerate}[$\bullet$]\setlength{\itemsep}{5pt}
\item
$\delta(\emptyset,\emptyset)=0$
\item
$\delta(F,\emptyset)=\delta(F-r_F,\emptyset)+\cdel(r_F)$
\item
$\delta(\emptyset,G)=\delta(\emptyset,G-r_G)+\cdel(r_G)$
\item
$\delta(F,G)=\min \left\{
\begin{array}{l}
\delta(F-r_F,G)+\cdel(r_F), \\
\delta(F,G-r_G)+\cdel(r_G),\vphantom{\Bigg(} \\
\delta(R_F-r_F,R_G-r_G) \\
\quad{+}\ \delta(F-R_F,G-R_G) \\
\quad{+}\ \cmatch(r_F,r_G)
\end{array}
\right.$
\end{enumerate}
\end{lemma}

The above lemma yields an $O(m^2n^2)$ dynamic program algorithm: If
we index the vertices of the forests $F$ and $G$ according to their
postorder traversal position, then entries in the dynamic program
table correspond to pairs $(F',G')$ of subforests $F'$ of $F$ and
$G'$ of $G$ where $F'$ contains vertices $\{i_1,\ldots,j_1\}$ and
$G'$ contains vertices $\{i_2,\ldots,j_2\}$ for some $1 \le i_1 \le
j_1 \le n$ and $1 \le i_2 \le j_2 \le m$.

However, as we will presently see, only
$O(\min\{n_\height,n_\leaves\} \cdot \min\{m_\height, m_\leaves\}
\cdot n m)$ different {\em relevant subproblems} are encountered by
the recursion computing $\delta(F,G)$. We calculate the number of
\emph{relevant subforests} of $F$ and $G$ independently, where a
forest $F'$ (respectively $G'$) is a relevant subforest of $F$
(respectively $G$) if it shows up in the computation of
$\delta(F,G)$. Clearly, multiplying the number of relevant
subforests of $F$ and of $G$ is an upper bound on the total number
of relevant subproblems.

We focus on counting the number of relevant subforests of $F$. The
count for $G$ is similar. First, notice that for every node $v \in
F$, $F_v-v$ is a relevant subproblem. This is because the recursion
allows us to delete the rightmost root of $F$ repeatedly until $v$
becomes the rightmost root; we then match $v$ (i.e., relabel it) and
get the desired relevant subforest. A more general claim is stated
and proved later on in Lemma \ref{Lemma:AllSubTrees}. We define
$\mr{keyroots}(F)=\{\textrm{the root of }F\}\cup \{v\in F \mid
    v\textrm{ has a left sibling}\}$.
Every relevant subforest of $F$ is a prefix (with respect to the
postorder indices) of $F_v-v$ for some node $v\in \mr{keyroots}(F)$.
If we define $\mr{cdepth}(v)$ to be the number of keyroot ancestors
of $v$, and $\mr{cdepth}(F)$ to be the maximum $\mr{cdepth}(v)$ over
all nodes $v\in F$, we get that the total number of relevant
subforest of $F$ is at most
\begin{align*}
  \sum_{\!\!\!\!v\in \mr{keyroots}(F)\!\!\!\!}|F_v|
  &= \sum_{v\in F}\mr{cdepth}(v)\\
  &\leq \sum_{v\in F}\mr{cdepth}(F)\\
  &= |F|\mr{cdepth}(F).
\end{align*}

This means that given two trees, $F$ and $G$, of sizes $n$ and $m$
we can compute $\delta(F,G)$ in $O(\mr{cdepth}(F)\mr{cdepth}(G)mn)$
time. Shasha and Zhang also proved that for any tree $T$ of size
$n$, $\mr{cdepth}(T)\leq \min\{n_\height,n_\leaves\}$, hence the
result. In the worst case, this algorithm runs in $O(m^2n^2)=O(n^4)$
time.

\subsection{Klein's Algorithm~\cite{Klein}}
\label{klein} Klein's algorithm is based on a recursion similar to
Lemma~\ref{shasha}. Again, we consider forests $F$ and $G$ of sizes
$|F| = n \geq |G| = m$. Now, however, instead of recursing always on
the rightmost roots of $F$ and $G$, we recurse on the leftmost roots
if $|L_F|\le|R_F|$ and on the rightmost roots otherwise. In other
words, the ``direction'' of the recursion is determined by the
(initially) larger of the two forests. We assume the number of
relevant subforests of $G$ is $O(m^2)$; we have already established
that this is an upper bound.

We next show that Klein's algorithm yields only $O(n\log n)$
relevant subforests of $F$. The analysis is based on a technique
called \emph{heavy path decomposition} introduced by Harel and
Tarjan~\cite{Tarjan84}. Briefly: we mark the root of $F$ as
\emph{light}. For each internal node $v \in F$, we pick one of $v$'s
children of maximum size and mark it as {\em heavy}, and we mark all
the other children of $v$ as \emph{light}. We define
$\mr{ldepth}(v)$ to be the number of light nodes that are ancestors
of $v$ in $F$, and $\mr{light}(F)$ as the set of all light nodes in
$F$. By~\cite{Tarjan84}, for any forest $F$ and vertex $v\in F$,
$\mr{ldepth}(v)\leq \log|F|+O(1)$. Note that every relevant
subforest of $F$ is obtained by some $i\leq |F_v|$ many consecutive
deletions from $F_v$ for some light node $v$. Therefore, the total
number of relevant subforests of $F$ is at most
\begin{align*}
  \sum_{\!\!\!\!v\in \mr{light}(F)\!\!\!\!}|F_v|
  &= \sum_{v\in F}\mr{ldepth}(v)\\
  &\leq \sum_{v\in F}(\log|F|+O(1))\\
  &= O(|F|\log|F|).
\end{align*}

Thus, we get an $O(m^2n\log n)=O(n^3\log n)$ algorithm for computing
$\delta(F,G)$.

\subsection{The Decomposition Strategy Framework \label{strategy}}
Both Klein's and Shasha and Zhang's algorithms are based on
Lemma~\ref{shasha}. The difference between them lies in the choice
of when to recurse on the rightmost roots and when on the leftmost
roots. The family of \emph{decomposition strategy} algorithms based
on this lemma was formalized by Dulucq and Touzet in~\cite{Touzet}.

\begin{definition}[Strategy]
Let $F$ and $G$ be two forests. A strategy is a mapping from pairs
$(F',G')$ of subforests of $F$ and $G$ to $\{\Left, \Right\}$.
\end{definition}

Each strategy is associated with a specific set of recursive calls
(or a dynamic program algorithm). The strategy of Shasha and Zhang's
algorithm is $S(F',G')=\Right$ for all $F',G'$. The strategy of
Klein's algorithm is $S(F',G')= \Left$ if $|L_{F'}|\le|R_{F'}|$, and
$S(F',G')=\Right$ otherwise. Notice that Shasha and Zhang's strategy
does not depend on the input trees, while Klein's strategy depends
only on the larger input tree. Dulucq and Touzet proved a lower
bound of $\Omega(mn\log m\log n)$ time for any strategy based
algorithm.

%% file: Algorithm.tex
In this section we present our algorithm for computing $\delta(F,G)$
given two trees $F$ and $G$ of sizes $|F| = n \ge |G| = m$. The
algorithm recursively uses Klein's strategy in a divide-and-conquer
manner to achieve $O(n m^2 (1 + \log\frac{n}{m})) = O(n^3)$ running
time in the worst case. The algorithm's space complexity is $O(nm)$.
We begin with the observation that Klein's strategy always
determines the direction of the recursion according to the
$F$-subforest, even in subproblems where the $F$-subforest is
smaller than the $G$-subforest. However, it is not straightforward
to change this since even if at some stage we decide to switch to
Klein's strategy based on the other forest, we must still make sure
that all subproblems previously encountered are entirely solved. At
first glance this seems like a real obstacle since apparently we
only add new subproblems to those that are already computed.

For clarity we describe the algorithm recursively. A dynamic
programming description and a proof of the $O(mn)$ space complexity
will appear in the full version of this paper.

\begin{figure}[h!]
\centerline{
\includegraphics[scale=0.8]
{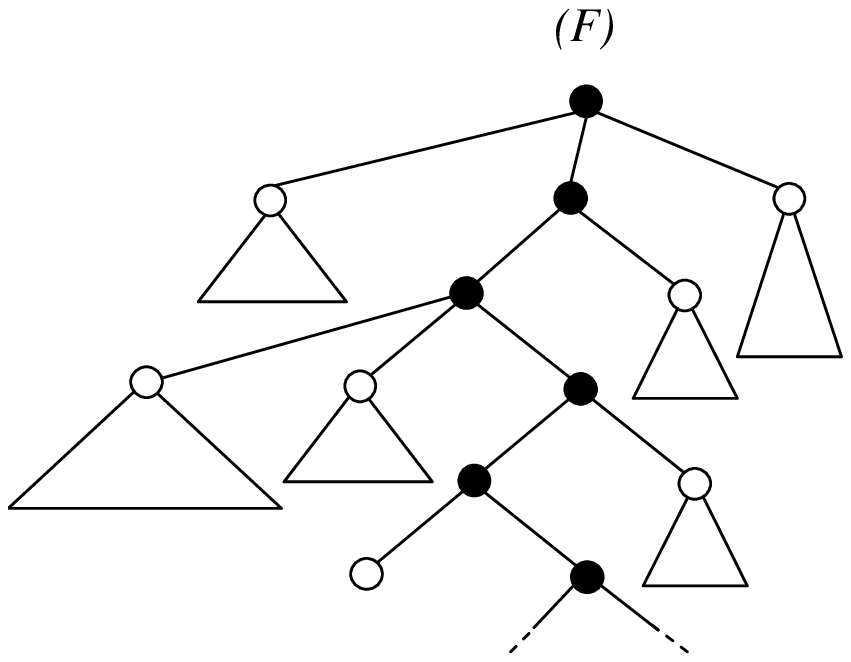} } \caption{A tree $F$ with $n$ nodes. The black
nodes belong to the heavy path. The white nodes are in
$\TopLight_F$, and the size of each subtree rooted at a white node
is at most $\frac{n}{2}$.} \label{LIGHT}
\end{figure}

For a tree $F$ of size $n$, define the set $\TopLight_F$ to be the
set of roots of the forest obtained by removing the heavy path of
$F$ (i.e., the unique path starting from the root along heavy
nodes). Note that $\TopLight_F$ is the set of light nodes with
$\ldepth$ 1 in $F$ (see the definition of $\ldepth$ in
section~\ref{klein}). This definition is illustrated in
Fig.~\ref{LIGHT}.
Note that the following two conditions are always satisfied:
\begin{itemize}
\item[\ \ ($\ast$)] $\displaystyle\sum_{v \in \TopLight_F} |F_v| \leq
n$.\medskip\\
This follows from the fact that $F_{v'}$ and $F_{v''}$ are disjoint
for any $v',v'' \in \TopLight_F$.\medskip
\item[\ \ ($\ast\ast$)] $|F_v| < \frac{n}{2}$
for every $v \in \TopLight_F$, since otherwise $v$ would be a heavy
node.
\end{itemize}

\paragraph{The Algorithm.} We compute $\delta(F,G)$ recursively as follows:

\begin{enumerate}[(1)]
\item If $|F|<|G|$, compute $\delta(G,F)$ instead. That is, we
order the pair $\{F,G\}$ such that $F$ is always the larger forest.
\item Recursively compute $\delta(F_v,G)$ for all $v \in \TopLight_F$.
Note that along the way this computes $\delta(F_{v'}-v',G_w-w)$ for
all $v'$ not in the heavy path of $F$ and for all $w \in G$.
\item Compute $\delta(F,G)$ using Klein's strategy
(matching and deleting either from the left or from the right
according to the larger of $F$ and $G$). Do not recurse into
subproblems that were previously computed in step (2).
\end{enumerate}

\noindent{The correctness of the algorithm follows immediately from
the correctness of Klein's algorithm. The algorithm is evidentally a
decomposition strategy algorithm, since for all subproblems, it
either deletes or matches the leftmost or rightmost roots.}

\paragraph{Time Complexity.} We show that our algorithm has a worst-case runtime of
$O(m^2n(1+\log\frac{n}{m}))= O(n^3)$.

We proceed by counting the number of subproblems computed in each
step of the algorithm. Let $R(F,G)$ denote the number of relevant
subproblems encountered by the algorithm in the course of computing
$\delta(F,G)$.

In step (2) we compute $\delta(F_v,G)$ for all $v \in \TopLight_F$.
Hence, the number of subproblems encountered in this step is
$\sum_{v \in \TopLight_F} R(F_v,G)$.

In step (3) we compute $\delta(F,G)$ using Klein's strategy. We
bound the number of relevant subproblems by multiplying the number
of relevant subforests in $F$ and in $G$. For $G$, we count all
possible $O(|G|^2)$ subforests obtained by left and right deletions.
Note that for any node $v'$ not in the heavy path of $F$, the
subproblem obtained by matching $v'$ with any node $w$ in $G$ was
already computed in step (2). This is because any such $v'$ is
contained in $F_v$ for some $v \in \TopLight_F$, so
$\delta(F_{v'}-v', G_w-w)$ is computed in the course of computing
$\delta(F_v,G)$ (we prove this formally in
Lemma~\ref{Lemma:AllSubTrees}). Furthermore, note that in Klein's
algorithm, a node $v$ on the heavy path of $F$ cannot be matched or
deleted until the remaining subforest of $F$ is precisely $F_v$. At
this point, both matching $v$ or deleting $v$ results in the same
new relevant subforest $F_v-v$. This means that we do not have to
consider matchings of nodes when counting the number of relevant
subproblems in step (3). It suffices to consider only the $|F|$
subforests obtained by deletions according to Klein's strategy.
Thus, the total number of new subproblems encountered step (3) is
bounded by $|G|^2|F|$.

We have established that $R(F,G)$ is at most
\begin{align*}
|G|^2|F| +  \sum_{v \in \TopLight_F} R(F_v,G)  \textrm{,  if } |F| \ge |G| \\
|F|^2|G| +  \sum_{w \in \TopLight_G} R(F,G_w)  \textrm{,  if } |F| <
|G|
\end{align*}

We first show, by a crude estimate, that this leads to an $O(n^3)$
runtime. Later, we analyze the dependency on $m$ and $n$ accurately.
\begin{lemma} $R(F,G) \leq 4(|F||G|)^{3/2}$. \label{Lemma:sqrt}
\end{lemma}
\begin{proof}
We proceed by induction on $|F|+|G|$. There are two symmetric cases.
If $|F| \ge |G|$ then $R(F,G) \le |G|^2|F| +  \sum_{v \in
\TopLight_F} R(F_v,G) $.
Hence, by the inductive assumption,
\begin{align*}
R(F,G) &\le  |G|^2|F| + \sum_{\!\!\!\!\!\!\!v \in
\TopLight_F\!\!\!\!\!\!\!} 4 (|F_v||G|)^{3/2} \\
& \le |G|^2|F| + 4|G|^{3/2} \sum_{\!\!\!\!\!\!\!v \in
\TopLight_F\!\!\!\!\!\!\!} |F_v|^{3/2} \\
& \le |G|^2|F| + \mbox{}\\
& {\phantom\le}\quad\ 4|G|^{3/2} \sum_{\!\!\!\!\!\!\!v \in
\TopLight_F\!\!\!\!\!\!\!} |F_v| \hspace*{-20pt}
\max_{\!\!\!\!\substack{\mbox{\phantom{-}} \\\ \ \ \ \ \ \ \
 v \in\TopLight_F}\!\!\!\!}\!\!\!\!\!\!\!\!\!\!\!\sqrt{|F_v|} \\
& \leq |G|^2|F| + 4 |G|^{3/2} |F| \sqrt{|F|/2} \\
&= |G|^2|F| + 2\sqrt{2} (|F||G|)^{3/2} \\
& \leq 4(|F||G|)^{3/2}.
\end{align*}
Here we have used facts $(\ast)$ and $(\ast\ast)$ and the fact that
$|F| \geq |G|$. The case where $|F| < |G|$ is symmetric.
 \qed
\end{proof}

This crude estimate gives a worst-case runtime of $O(n^3)$. We now
analyze the dependence on $m$ and $n$ more accurately. Along the
recursion defining the algorithm, we view step (2) as only making
recursive calls, but not producing any relevant subproblems. Rather,
every new relevant subproblem is created in step (3) for a unique
recursive call of the algorithm. So when we count relevant
subproblems, we sum the number of new relevant subproblems
encountered in step (3) over all recursive calls to the algorithm.

We define sets $A,B \subseteq F$ as follows:
\begin{align*}
  A &= \big\{a \in \light(F) : |F_a| \ge m\big\}\\
  B &= \big\{b \in F{-}A : b \in \TopLight_{F_a}\text{ for some }a\in A\big\}.
\end{align*}
Note that the root of $F$ belongs to $A$.
We count separately:
\begin{enumerate}[\ (i)]
  \item
    the relevant subproblems created in just step (3) of recursive
    calls $\delta(F_a,G)$ for all $a \in A$, and
  \item
    the relevant subproblems encountered in the entire computation
    of $\delta(F_b,G)$ for all $b \in B$ (i.e., $\sum_{b \in
    B}R(F_b,G)$).
\end{enumerate}
Together, this counts all relevant subproblems for the original
$\delta(F,G)$. To see this, consider the original call
$\delta(F,G)$. Certainly, the root of $F$ is in $A$. So all
subproblems generated in step (3) of $\delta(F,G)$ are counted in
(i). Now consider the recursive calls made in step (2) of
$\delta(F,G)$. These are precisely $\delta(F_v,G)$ for $v \in
\TopLight_F$. For each $v \in \TopLight_F$, notice that $v$ is
either in $A$ or in $B$; it is in $A$ if $|F_v| \ge m$, and in $B$
otherwise. If $v$ is in $B$, then all subproblems arising in the
entire computation of $\delta(F_v,G)$ are counted in (ii). On the
other hand, if $v$ is in $A$, then we are in analogous situation
with respect to $\delta(F_v,G)$ as we were in when we considered
$\delta(F,G)$ (i.e., we count separately the subproblems created in
step (3) of $\delta(F_v,G)$ and the subproblems coming from
$\delta(F_u,G)$ for $u \in \TopLight_{F_v}$).

Earlier in this section, we saw that the number of subproblems
created in step (3) of $\delta(F,G)$ is $|G|^2|F|$. In fact, for any
$a \in A$, by the same argument, the number of subproblems created
in step (3) of $\delta(F_a,G)$ is $|G|^2|F_a|$. Therefore, the total
number of relevant subproblems of type (i) is $|G|^2\sum_{a \in
A}|F_a|$. For $v \in F$, define $\mr{depth}_A(v)$ to be the number
of ancestors of $v$ that lie in the set $A$. We claim that
$\mr{depth}_A(v) \le 1+\log\frac n m$ for all $v \in F$. To see
this, consider any sequence $a_0,\dots,a_k$ in $A$ where $a_i$ is a
descendent of $a_{i-1}$ for all $i \in [1,k]$. Note that $|F_{a_i}|
\le \frac 1 2|F_{a_{i-1}}|$ for all $i \in [1,k]$ since the $a_i$
are light nodes, and note that $|F_{a_k}| \ge m$ by the definition
of $A$. It follows that $k \le \log\frac n m$, i.e., $A$ contains no
sequence of descendants of length $> 1 + \log \frac n m$. So clearly
every $v \in F$ has $\mr{depth}_A(v) \le 1 +\log\frac n m$.

We now have the number of relevant subproblems of type (i) as
\begin{align*}
  |G|^2\sum_{a \in A}|F_a|
  &= m^2\sum_{v\in F}\mr{depth}_A(v)\\
  &\leq m^2\sum_{v\in F}(1+\log \frac{n}{m})\\
  &=m^2n(1+\log\frac{n}{m}).
\end{align*}

The relevant subproblems of type (ii) are counted by $\sum_{b \in B}
R(F_b,G)$. Using Lemma~\ref{Lemma:sqrt}, we have
\begin{align*}
  \sum_{b \in B}R(F_b,G) &\le 4|G|^{3/2}\sum_{b \in B}|F_b|^{3/2}\\
  &\le 4|G|^{3/2}\sum_{b \in B}|F_b|\max_{b\in B}\sqrt{|F_b|}\\
  &\le 4|G|^{3/2}|F|\sqrt m = 4m^2n.
\end{align*}
Here we have used the facts that $|F_b| < m$ and $\sum_{b \in
B}|F_b| \le |F|$ (since the trees $F_b$ are disjoint for different
$b \in B$). Therefore, the total number of relevant subproblems for
$\delta(F,G)$--and hence the runtime of the algorithm--is at most
$m^2n(1+\log\frac{n}{m}) + 4m^2n = O(m^2n(1+\log\frac n m))$.

\paragraph{Unrooted Trees.} Our algorithm can be adapted
to compute edit distance of \emph{unrooted ordered trees}. An
unrooted ordered tree is an acyclic graph with a cyclic ordering
defined on the edges incident on each node in the graph. In the
modified algorithm, we arbitrarily choose a root for the larger of
the two trees. We change the first recursive level of the algorithm,
so that it now computes the edit distance with respect to any
possible choice of a root for the smaller tree. This does not change
the time complexity since the number of different relevant
subforests for a tree of size $m$ is bounded by $m^2$ whether we
consider a single choice for the root or all possible choices. This
idea will be described in detail in the full version of this paper.

%% file: LowerBound.tex

In this section we present a lower bound on the worst-case runtime
of strategy algorithms. We first give a simple proof of an
$\Omega(m^2n)$ lower bound. In the case where $m = \Theta(n)$, this
gives a lower bound of $\Omega(n^3)$ which shows that our algorithm
is worst-case optimal among all strategy-based algorithms. To prove
that our algorithm is worst-case optimal for any $m \leq n$, we
analyze a more complicated scenario that gives a lower bound of
$\Omega(m^2n(1+\log{\frac{n}{m}}))$, matching the running time of
our algorithm.

In analyzing strategies we will use the notion of a
\emph{computational path}, which corresponds to a specific sequence
of recursion calls. Recall that for all subforest-pairs $(F',G')$,
the strategy $S$ determines a direction: either $\Right$ or $\Left$.
The recursion can either delete from $F'$ or from $G'$ or match. A
computational path is the sequence of operations taken according to
the strategy in a specific sequence of recursive calls. For
convenience, we sometimes describe a computational path by the
sequence of subproblems it induces, and sometimes by the actual
sequence of operations: either ``delete from the $F$-subforest'',
``delete from the $G$-subforest'', or ``match''.

The following lemma states that every strategy computes the edit
distance between every two root-deleted subtrees of $F$ and $G$.

\begin{lemma}
\label{Lemma:AllSubTrees}%
For any strategy $S$, the pair $(F_v{-}v,G_w{-}w)$ is a relevant
subproblem for all $v \in F$ and $w \in G$.
\end{lemma}
\begin{proof}
First note that a node $v' \in F_v$ (respectively, $w' \in G_w$) is
never deleted or matched before $v$ (respectively, $w$) is deleted
or matched. Consider the following computational path:
\begin{itemize}
  \item
    Delete from $F$ until $v$ is either the leftmost or the rightmost
    root.
  \item
    Next, delete from $G$ until $w$ is either the leftmost or the
    rightmost root.
\end{itemize}
Let $(F',G')$ denote the resulting subproblem. There are four cases
to consider.
\begin{enumerate}\setlength{\itemsep}{4pt}
  \item
    {\sl $v$ and $w$ are the rightmost \tu(leftmost\tu)
    roots of $F'$ and $G'$, and $S(F',G')=\Right$
    \tu($\Left$\tu).}\medskip

    Match $v$ and $w$ to get the desired
    subproblem.
  \item
    {\sl $v$ and $w$ are the rightmost \tu(leftmost\tu) roots of $F'$ and
    $G'$, and $S(F',G')=\Left$ \tu($\Right$\tu).}\medskip

    Note that at least one of $F',G'$ is not a tree (since otherwise
    this is case (1)). Delete from one which is not a tree. After a
    finite number of such deletions we have reduced to case (1),
    either because $S$ changes direction, or because both forests
    become trees whose roots are $v,w$.
  \item
    {\sl $v$ is the rightmost root of $F'$, $w$ is the leftmost root
    of $G'$.}\medskip

    If $S(F',G')=\Left$, delete from $F'$; otherwise delete from
    $G'$. After a finite number of such deletions this reduces to
    one of the previous cases when one of the forests becomes a
    tree.
  \item
    {\sl $v$ is the leftmost root of $F'$, $w$ is the rightmost root of
    $G'$.}\medskip

    This case is symmetric to (3).\qed
\end{enumerate}
\end{proof}
We now turn to the $\Omega(m^2n)$ lower bound on the number of
relevant subproblems for any strategy.
\begin{lemma}
\label{Lemma:LowerBound}%
For any strategy $S$, there exists a pair of trees $(F,G)$ with
sizes $n,m$ respectively, such that the number of relevant
subproblems is $\Omega(m^2n)$.
\end{lemma}
\begin{figure}[h!]
\begin{center}
\includegraphics[scale=0.4]{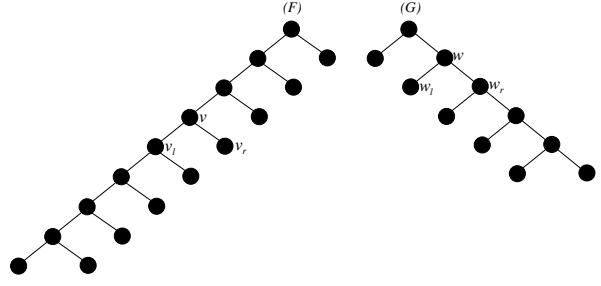}
\caption {The two trees used to prove an $\Omega(m^2n)$ lower
bound.}
\end{center} \label{combs}
\end{figure}

\begin{proof}
Let $S$ be an arbitrary strategy, and consider the trees $F$ and $G$
depicted in Fig.~\ref{combs}. According to lemma
\ref{Lemma:AllSubTrees}, every pair $(F_v{-}v,G_w{-}w)$ where $v \in
F$ and $w \in G$ is a relevant subproblem for $S$. Focus on such a
subproblem where $v$ and $w$ are internal nodes of $F$ and $G$.
Denote $v$'s right child by $v_r$ and $w$'s left child by $w_\ell$.
Note that $F_v{-}v$ is a forest whose rightmost root is the node
$v_r$. Similarly, $F_w{-}w$ is a forest whose leftmost root is
$w_\ell$. Starting from $(F_v{-}v,G_w{-}w)$, consider the
computational path $c_{v,w}$ that deletes from $F$ whenever the
strategy says $\Left$ and deletes from $G$ otherwise. In both cases,
neither $v_r$ nor $w_\ell$ is deleted. Such deletions can be carried
out so long as both forests are non-empty.

The length of this computational path is at least
$\min\{|F_v|,|G_w|\}-1$. Note that for each subproblem $(F',G')$
along this computational path, $v_r$ is the rightmost root of $F'$
and $w_\ell$ is the leftmost root of $G'$. It follows that for every
two distinct pairs $(v_1,w_1) \ne (v_2,w_2)$ of internal nodes in
$F$ and $G$, the relevant subproblems occurring along the
computational paths $c_{v_1,w_1}$ and $c_{v_2,w_2}$ are disjoint.
Since there are $\frac{n}{2}$ and $\frac{m}{2}$ internal nodes in
$F$ and $G$ respectively, the total number of subproblems along the
$c_{v,w}$ computational paths is given by:
%
\begin{align*}
\sum_{(v,w)\textrm{ internal nodes}} \min\{|F_v|,|G_w|\}-1 &= \\
\sum_{i=1}^{\frac{n}{2}} \sum_{j=1}^{\frac{m}{2}} \min\{2i,2j\} &=
\Omega(m^2n)
\end{align*}
\qed
\end{proof}

The $\Omega(m^2n)$ lower bound established by
Lemma~\ref{Lemma:LowerBound} is tight if $m = \Theta(n)$, since in
this case our algorithm achieves an $O(n^3)$ runtime.


To establish a tight bound when $m$ is not $\Theta(n)$, we use the
following technique for counting relevant subproblems. We associate
a subproblem consisting of subforests $(F',G')$ with the unique pair
of vertices $(v,w)$ such that $F_v,G_w$ are the smallest trees
containing $F',G'$ respectively. For example, for nodes $v$ and $w$
with at least two children, the subproblem $(F_v{-}v,G_w{-}w)$ is
associated with the pair $(v,w)$. Note that all subproblems
encountered in a computational path starting from
$(F_v{-}v,G_w{-}w)$ until the point where either forest becomes a
tree are also associated with $(v,w)$.

\begin{figure}[h!]
\begin{center}
\includegraphics[scale=0.52]{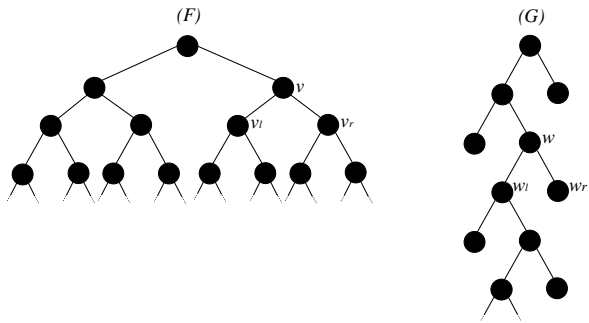}
\caption {The two trees used to prove $\Omega(m^2n\log\frac{n}{m})$
lower bound.} \label{zigzag}
\end{center}
\end{figure}

\begin{lemma}
\label{Lemma:TightBound}%
For every strategy $S$, there exists a pair of trees $(F,G)$ with
sizes $n \ge m$ such that the number of relevant subproblems is
$\Omega(m^2n\log\frac{n}{m})$.
\end{lemma}

\begin{proof}
Consider the trees illustrated in Fig.~\ref{zigzag}. The $n$-sized
tree $F$ is a complete balanced binary tree, and $G$ is a ``zigzag''
tree of size $m$. Let $w$ be an internal node of $G$ with a single
node $w_r$ as its right subtree and $w_\ell$ as a left child. Denote
$m' = |G_w|$. Let $v$ be a node be a node in $F$ such that $F_v$ is
a tree of size $n'+1$ such that $n' \geq 4m \geq 4m'$. Denote $v$'s
left and right children $v_\ell$ and $v_r$ respectively. Note that
$|F_{v_\ell}| = |F_{v_r}| = \frac{n'}{2}$

Let $S$ be an arbitrary strategy. We aim to show that the total
number of relevant subproblems associated with $(v,w)$ or with
$(v,w_\ell)$ is at least $\frac{n'm'}{4}$. Let $c$ be the
computational path that always deletes from $F$ (no matter whether
$S$ says $\Left$ or $\Right$). We consider two complementary
cases.\bigskip


\noindent\textsc{Case 1:}\ $\frac{n'}4$ left deletions occur in the
computational path $c$, and at the time of the $\frac{n'}4$th left
deletion, there were fewer than $\frac{n'}4$ right
deletions.\bigskip


We define a set of new computational paths $\{c_j\}_{1\le j \le
\frac{n'}4}$ where $c_j$ deletes from $F$ up through the $j$th left
deletion, and thereafter deletes from $F$ whenever $S$ says $\Right$
and from $G$ whenever $S$ says $\Left$. At the time the $j$th left
deletion occurs, at least $\frac{n'}{4} \ge m'-2$ nodes remain in
$F_{v_r}$ and all $m'-2$ nodes are present in $G_{w_\ell}$. So on
the next $m'-2$ steps along $c_j$, neither of the subtrees $F_{v_r}$
and $G_{w_\ell}$ is totally deleted. Thus, we get $m'-2$ distinct
relevant subproblems associated with $(v,w)$. Notice that in each of
these subproblems, the subtree $F_{v_\ell}$ is missing exactly $j$
nodes. So we see that, for different values of $j \in
[1,\frac{n'}4]$, we get disjoint sets of $m'-2$ relevant
subproblems. Summing over all $j$, we get $\frac{n'}{4}(m'-2)$
distinct relevant subproblems associated with $(v,w)$.\bigskip


\noindent\textsc{Case 2:}\ $\frac{n'}4$ right deletions occur in the
computational path $c$, and at the time of the $\frac{n'}4$th right
deletion, there were fewer than $\frac{n'}4$ left deletions.\bigskip

We define a different set of computational paths $\{\gamma_j\}_{1\le
j \le \frac{n'}4}$ where $\gamma_j$ deletes from $F$ up through the
$j$th right deletion, and thereafter deletes from $F$ whenever $S$
says $\Left$ and from $G$ whenever $S$ says $\Right$ (i.e.,
$\gamma_j$ is $c_j$ with the roles of $\Left$ and $\Right$
exchanged). Similarly as in case 1, for each $j\in[1,\frac{n'}4]$ we
get $m'-2$ distinct relevant subproblems in which $F_{v_r}$ is
missing exactly $j$ nodes. All together, this gives
$\frac{n'}4(m'-2)$ distinct subproblems. Note that since we never
make left deletions from $G$, the left child of $w_\ell$ is present
in all of these subproblems. Hence, each subproblem is associated
with either $(v,w)$ or $(v,w_\ell)$.\bigskip

In either case, we get $\frac{n'}4(m'-2)$ distinct relevant
subproblems associated with $(v,w)$ or $(v,w_\ell)$. To get a lower
bound on the number of problems we sum over all pairs $(v,w)$ with
$G_w$ being a tree whose right subtree is a single node, and $|F_v|
\geq 4m$. There are $\frac{m}{4}$ choices for $w$ corresponding to
tree sizes $4j$ for $j\in[1,\frac{m}{4}]$. For $v$, we consider all
nodes of $F$ whose distance from a leaf is at least $\log(4m)$. For
each such pair we count the subproblems associated with $(v,w)$ and
$(v,w_\ell)$. So the total number of relevant subproblems counted in
this way is
\begin{align*}
\sum_{v,w} \frac{|F_v|}{4}(|G_w|-2) &=
\frac 1 4\sum_v |F_v|\sum_{j=1}^{\frac{m}{4}} (4j-2) \\
&= \frac 1 4\sum_{i=\log4m}^{\log n} \frac{n}{2^i}{\cdot}2^i \sum_{j=1}^{\frac{m}{4}} (4j-2) \\
&= \Omega(m^2n \log \frac{n}{m})
\end{align*}
\qed
\end{proof}

\begin{lemma}
For every strategy $S$ and $n \ge m$, there exist trees $F$ and $G$
of sizes $\Theta(n)$ and $\Theta(m)$ for which $S$ has
$\Omega(m^2n(1+\log \frac{n}{m}))$ relevant subproblems.
\end{lemma}
\begin{proof}
If $m=\Theta(n)$ then this bound is $\Omega(m^2n)$ as shown in Lemma
\ref{Lemma:LowerBound}. Otherwise, this bound is $\Omega(m^2n \log
\frac{n}{m})$ which was shown in Lemma \ref{Lemma:TightBound}. \qed
\end{proof}

%% file: Conclusions.tex

We presented a new $O(n^3)$-time and $O(n^2)$-space algorithm for
computing the tree edit distance between two ordered trees. Our
algorithm is not only faster than all previous algorithms in the
worst case, but we have proved it is optimal within the broad class
of decomposition strategy algorithms. As a consequence, any future
improvements in terms of worst-case time complexity would have to
find an entirely new approach. We obtain similar results when
considering the sizes $m$ and $n$ of the input trees as separate
parameters.

The novelty of our dynamic program is that it is both symmetric in
its two inputs as well as adaptively dependant on them. This general
notion may also be applied in other scenarios where the known
dynamic programming solutions possess an inherent asymmetry.

The full version of the paper includes an explicit dynamic program
for our algorithm, a proof of the $O(n^2)$-space complexity, and an
adaptation of the algorithm for the edit distance problem on
unrooted trees.